# Modulation instability in nonlinear complex parity-time (PT) symmetric periodic structures

Amarendra K. Sarma*
*Department of Physics, Indian Institute of Technology Guwahati, Guwahati-781039, Assam, India*
*\* Corresponding author: aksarma@iitg.ernet.in*

We carry out a modulation instability (MI) analysis in nonlinear complex parity-time (PT) symmetric periodic structures. All the three regimes defined by the PT-symmetry breaking point or threshold, namely, below threshold, at threshold and above threshold are discussed. It is found that MI exists even beyond the PT-symmetry threshold indicating the possible existence of solitons or solitary waves, in conformity with some recent reports. We find that MI does not exist at the PT-symmetry breaking point in the case of normal dispersion below a certain nonlinear threshold. However, in the case of anomalous dispersion regime, MI does exist even at the PT-symmetry breaking point.

## 1. Introduction

Non-Hermitian parity-time (PT) symmetric quantum mechanics has been studied theoretically for a decade or so [1-5]. On the other hand, recently this concept of parity-time symmetry is realized experimentally in optics in the context of optical coupled waveguide system [6], optical mesh lattices [7] and metamaterials [8]. In passing, it may be noted that PT-symmetry finds experimental demonstration even in LRC circuit [9]. An optical system is said to exhibit PT-symmetry if the complex refractive-index distribution, $n(x) = n_R(x) + i\, n_I(x)$, of the system satisfies the condition: $n_R(x) = n_R(-x)$ and $n_I(x) = -n_I(-x)$. This physically implies that the refractive-index distribution must be an even function of position while the gain/loss profile must be odd. For a comprehensive review on the subject readers are referred to Ref. [10]. PT-symmetry in optics has become a topic of considerable interest in various contexts. To mention a few: beam dynamics in PT-symmetric structures have been investigated to show the possibility of double refraction, power oscillations and secondary emissions [11]. Another interesting development is related to the induction of unidirectional invisibility by PT-symmetric structures at the so-called PT-symmetric breaking point or threshold [12]. Again, PT-symmetric optical systems have found important applications in chip-scale optical isolators [13-14], laser amplifiers [15] and coherent perfect laser absorbers [16] etc. In the context of nonlinear optics, optical solitons in PT-synthetic lattices are investigated and found to be stable over a wide range of potential parameters [17]. The swing behavior of spatial solitons propagating along waveguides whose refractive indices in the transverse direction are perturbed by parity–time symmetric profiles is also investigated [18].

Recently, it has been demonstrated that slow Bragg solitons are possible in nonlinear PT-symmetric periodic structures [19]. It is shown that the grating band structure is modified owing to the PT-symmetric component of the periodic optical refractive index. In the linear regime of such structures, for some range of the gain-loss coefficient, there exists a threshold value above which the system no longer exhibits PT-symmetry while below this point the system maintains its PT-symmetric characteristics. In this work, we intend to carry out modulation instability (MI) analysis in the above mentioned system. It is well known that modulation instability (MI) is a fundamental and ubiquitous process that appears in most nonlinear systems in nature [20-21]. MI refers to the exponential growth of a periodic modulation on a plane wave background. It occurs as a result of the interplay between nonlinearity and dispersion in the time domain or diffraction in the spatial domain. In the context of an optical fiber, modulation instability manifests in the temporal domain through the breakup of quasi-continuous wave (CW) radiation into a train of ultra-short pulses and has been explored as a possible means to generate pulse trains with high repetition rate. On the other hand, in the spectral domain MI corresponds to degenerate four-wave mixing (FWM) phase-matched through self-phase modulation [21]. MI is observed in numerous physical situations including water waves, plasma waves, laser beams, and electromagnetic transmission lines [22-29]. Modulation instability in optical fiber is studied quite extensively [21,30-31]. It is studied in the context of the negative index material also [22, 26]. Experimental and theoretical investigation towards MI in Bragg gratings and periodic structures is also carried out by many authors [32-34]. In a recent study, traveling wave solution in a nonlinear PT-symmetric Bragg grating structure is reported. It is demonstrated that below PT-threshold, there exists a bright solitary-wave solution for forward waves and a dark solitary-wave solution for backward waves [35]. It is remarkable to note that depending on some suitable choice of the parameter values of the system, there may even exist the so-called optical Rogue waves [35]. However, we note that though various solitary waves have been predicted in nonlinear PT-symmetric periodic structures, no study is carried out, to the best of our knowledge, to study the phenomenon of modulation instability, which is considered as a precursor to soliton formation. Moreover, MI analysis may throw new insights into the working of such new novel nonlinear optical systems. It should be noted that the so-called Bragg soliton is intrinsically related to MI in periodic structures and have been studied quite extensively [34, 36-38]. In this work we investigate MI in nonlinear complex parity-time (PT) symmetric periodic structures in all the three regimes: below PT-symmetry threshold, at threshold and above threshold. It is worthwhile to mention that our analysis of MI in PT-symmetric periodic structures takes a similar procedure found in the case of MIs in Bragg gratings [34]. The rest of the paper is organized as follows: In Sec. 2 we present the theoretical model and linear stability analysis. Sec. 3 contains results and discussions followed by conclusions in Sec. 4.

## 2.Theoretical Model

The coupled wave equations in nonlinear PT-symmetric periodic potentials have been derived by M. A. Miri et al. [17]:

$$i\left(\frac{\partial E_f}{\partial z} + \frac{1}{v_g}\frac{\partial E_f}{\partial t}\right) + \delta_0 E_f + (k_0 + g_0)E_b + \gamma_0\left(|E_f|^2 + 2|E_b|^2\right)E_f = 0 \quad (1a)$$

$$i\left(\frac{\partial E_b}{\partial z} - \frac{1}{v_g}\frac{\partial E_b}{\partial t}\right) - \delta_0 E_b - (k_0 - g_0)E_f - \gamma_0\left(2|E_f|^2 + |E_b|^2\right)E_b = 0 \quad (1b)$$

They have considered an optical fiber with the core refractive index profile given by $n = n_0 + n_{1R}\cos(2\pi z/\Lambda) + in_{1I}\sin(2\pi z/\Lambda) + n_2|E|^2$, where $n_0$ is the refractive index of the background material, $n_{1R}$ and $n_{1I}$ are the real and the imaginary part of the perturbed complex refractive index distribution. $n_2$ is the so-called nonlinear Kerr parameter and $\Lambda$ is the grating period. In Eq. (1), $E_f(z,t)$ and $E_b(z,t)$ represent the slowly varying amplitudes for the forward and backward waves, respectively. $v_g = c/n_0$ is the velocity in the background material, $k_0 = \pi n_{1R}/\lambda_0$ is the coupling coefficient arising from the real Bragg grating itself, $g_0 = \pi n_{1I}/\lambda_0$ is the anti-symmetric coupling coefficient arising from the complex PT-potential term. On the other hand, $\delta_0 = (\omega_0 - \omega_B)/v_g$ is the measure of detuning from the Bragg frequency and $\gamma_0$ is the self-phase modulation parameter.

In order to simplify our analysis, we adopt the following dimensionless units:
$$\xi = z/z_0, \tilde{E}_f = E_f/\sqrt{P_0}, \tilde{E}_b = E_b/\sqrt{P_0}, \tau = t/T_0 \quad (2)$$
Here $T_0$ and $P_0$ are respectively, the pulse width and peak power of the laser radiation. The length scale $z_0$ is defined such that, $z_0 = v_g T_0$. Using these units, Eq. (1) could be rewritten as follows:

$$i\left(\frac{\partial \tilde{E}_f}{\partial \xi} + \frac{\partial \tilde{E}_f}{\partial \tau}\right) + \delta\tilde{E}_f + (k+g)\tilde{E}_b + \gamma\left(|\tilde{E}_f|^2 + 2|\tilde{E}_b|^2\right)\tilde{E}_f = 0 \quad (3a)$$

$$i\left(\frac{\partial \tilde{E}_b}{\partial \xi} - \frac{\partial \tilde{E}_b}{\partial \tau}\right) - \delta\tilde{E}_b - (k-g)\tilde{E}_f - \gamma\left(2|\tilde{E}_f|^2 + |\tilde{E}_b|^2\right)\tilde{E}_b = 0 \quad (3b)$$

Here $\delta = \delta_0 z_0$, $k = k_0 z_0$, $g = g_0 z_0$ and $\gamma = \gamma_0 P_0 z_0$.
In the continuous wave limit, we now assume the following forms for the solutions of Eqs. (1):
$$\tilde{E}_f = A_f e^{iq\xi} \text{ and } \tilde{E}_b = A_b e^{iq\xi} \quad (4)$$
Here $A_f^2 + A_b^2 = 1$, such that $|E_f|^2 + |E_b|^2 = P_0$. Introducing the parameter $f = A_b/A_f$,
we write: $A_f = \sqrt{1/(1+f^2)}$ and $A_b = \sqrt{1/(1+f^2)}\,f$,
where $f$ represents the ratio of the input amplitudes of the backward and the forward propagating waves. Putting (4) in (3), we can easily derive the following nonlinear dispersion relations:

$$\delta = -\frac{k(1+f^2)-g(1-f^2)}{2f} - \frac{3\gamma}{2}, q == -\frac{k(1-f^2)-g(1+f^2)}{2f} - \frac{\gamma}{2}\left(\frac{1-f^2}{1+f^2}\right) \quad (5)$$

In the absence of nonlinear effects, one can easily obtain the dispersion relation: $q^2 = \delta^2 + (g^2 - k^2)$. Fig. 1 depicts the dispersion curves for both the passive medium ($g = 0$) and the PT-symmetric medium. It is important to note that the PT-symmetry threshold of the given system is defined by: $g = k$ [19]. The dispersion curves are divided into two regimes, namely, the normal dispersion (lower branch) and the anomalous dispersion (upper branch) depending to the sign of $f$ [33-34]. $f > 0$ corresponds to the normal dispersion, while $f < 0$ refers to the anomalous dispersion regimes. Fig. 1(a) depicts the usual results already known for normal periodic structures [34]. On the other hand, Fig. 1(b) seems to offer something interesting. One should note the distinct features of the dispersion curves in all the three regimes of PT-symmetry: below threshold, at threshold and above threshold, explained in the figure captions.

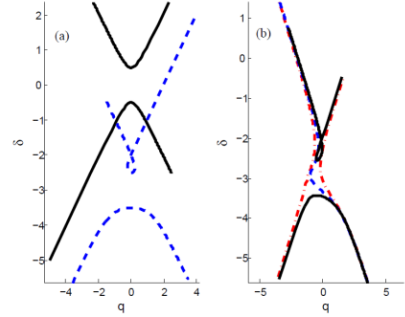

Fig. 1. (Color online) Dispersion curves traced by $q$ and $\delta$ with $k = 0.5$. Upper branch refers to $f < 0$ (anomalous dispersion) and lower branch refers to $f > 0$ (normal dispersion). (a) case of passive medium with $g = 0$ ; solid curves represent dispersion curves without nonlinearity, while the dashed one with nonlinearity, $\gamma = 2$. (b) case of PT-symmetric medium with $\gamma = 2$. Solid curve refers to below threshold ($g = 0.25$ ), dashed curve refers to at threshold ($g = 0.5$ ) while dash-dotted one refers to above threshold case ($g = 0.75$ ).

The dispersion curve for a passive medium with and without nonlinearity is easy to understand and well explained in Ref.[34]. Fig. 1(b) depicts the dispersion curves for a PT-symmetric nonlinear medium in various regimes. It is clear that the presence of the PT-symmetric term effectively shift the photonic band gap. On the other hand, nonlinearity leads to the formation of a loop on the upper branch (where $f < 0$) beyond a critical power level below the PT-symmetric threshold. It is interesting to note, however, is that this loop disappears at and above the PT-symmetric threshold. Also, the so-called photonic band gap disappears at and above the PT-symmetric threshold.

Now, in order to derive the MI gain spectrum we follow the standard procedure [33]. The steady state solutions (4) are perturbed slightly such that
$$u = (u_0 + \varepsilon_1)\exp(iq\xi)\exp(-i\delta\xi) \quad (6a)$$
$$v = (v_0 + \varepsilon_2)\exp(iq\xi)\exp(i\delta\xi) \quad (6b)$$
with $|\varepsilon_1, \varepsilon_2| \ll u_0, v_0$. Putting (6) in (3) and after linearization we obtain:

$$i\left(\frac{\partial \varepsilon_1}{\partial \xi} + \frac{\partial \varepsilon_1}{\partial \tau}\right) - (q-\delta)\varepsilon_1 + (k+g)\varepsilon_2 + \gamma u_0^2(2\varepsilon_1 + \varepsilon_1^*) + 2\gamma u_0 v_0(\varepsilon_2 + \varepsilon_2^*) + 2\gamma v_0^2 \varepsilon_1 = 0 \quad (7a)$$

$$i\left(\frac{\partial \varepsilon_2}{\partial \xi} - \frac{\partial \varepsilon_2}{\partial \tau}\right) - (q+\delta)\varepsilon_1 - (k-g)\varepsilon_1 - \gamma v_0^2(2\varepsilon_2 + \varepsilon_2^*) - 2\gamma u_0 v_0(\varepsilon_1 + \varepsilon_1^*) - 2\gamma u_0^2 \varepsilon_2 = 0 \quad (7b)$$

In order to solve the above set of coupled equations, we assume a plane wave ansatz, constituted of both forward and backward propagation, having the following form:

$$\varepsilon_1 = a_1 exp[i(K\xi - \Omega\tau)] + a_2 exp[-i(K\xi - \Omega\tau)] \quad (8a)$$
$$\varepsilon_2 = b_1 exp[i(K\xi - \Omega\tau)] + b_2 exp[-i(K\xi - \Omega\tau)] \quad (8b)$$

Here $a_j$ and $b_j$ are real constants with $j = 1,2$. $K$ and $\Omega$ are the normalized propagation constant and the perturbation frequency respectively. Putting (8) in (7) we obtain a set of four linearly coupled equations for $a_j$ and $b_j$, which has a nontrivial solution only when the $4 \times 4$ determinant formed by the coefficients matrix vanishes. This results in a fourth order polynomial in $\Omega$ :

$$\Omega^4 - \alpha\Omega^2 + \beta\Omega + \eta = 0 \quad (9)$$

where $\alpha = 2K^2 + AC + FH + BG - DE$, $\beta = 2K(FH - AC)$ and
$\eta = K^4 + (BG - DE - AC - FH)K^2 + ACFH$
$+ ADFG - BCEH - BDEG$, with
$A = -(q - \delta) + 3\gamma u_0^2 + 2\gamma v_0^2$, $B = k + g + 4\gamma u_0 v_0$, $C = -(q - \delta) + \gamma u_0^2 + 2\gamma v_0^2$, $D = k + g$, $E = -k + g - 4\gamma u_0 v_0$, $F = -(q + \delta) - 3\gamma v_0^2 - 2\gamma u_0^2$, $G = k - g$, and $H = -(q + \delta) - \gamma v_0^2 - 2\gamma u_0^2$.

The four roots of the polynomial can be written as

$$\Omega = \pm\frac{1}{2}\sqrt{p} \pm \frac{1}{2}\sqrt{2\alpha - p - 2\beta/\sqrt{p}}$$

where $q = -2\alpha^3 + 27\beta^2 + 72\alpha\eta$, $p = \frac{2}{3}\alpha + (\alpha^2 + 12\eta)/(3r) + \frac{1}{3}r$ and $r = \left[q + \sqrt{-4(\alpha^2 + 12\eta)^3 + q^2}\right]^{\frac{1}{3}}/2^{\frac{1}{3}}$

The MI phenomenon is measured by a gain given by $G = |Im \Omega_M|$, where $Im \Omega_M$ denotes the imaginary part of $\Omega_M$, where $\Omega_M$ is the root with the largest imaginary part.

### 3. Modulation Instability Analysis

In this section, we discuss the gain spectrum of the modulation instability based on the roots of the polynomial (9). Fig. 2 depicts the gain spectrum for three different regimes defined by the PT-symmetry threshold. It can be seen that the behavior of the gain spectrums are distinctly different in both the normal and the anomalous dispersion regimes. At the PT-symmetry threshold point, $g = k$, the system exhibits no MI gain at all in the case of normal dispersion, but in the anomalous dispersion, the system does show MI gain. Again, the MI gain spectrum is symmetric around $K = 0$ above the PT symmetry threshold ($g > k$) for normal dispersion regime, but not so for other cases.

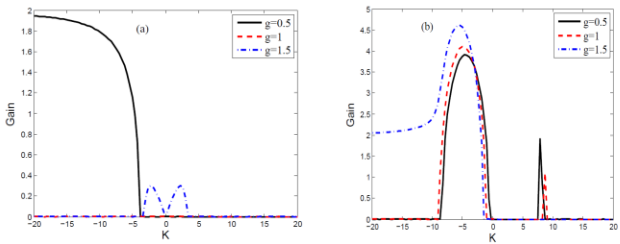

Fig. 2. (Color online) Gain spectrum for different regimes: below PT threshold ($g = 0.5$ );at threshold ($g = 1$ ) and above threshold ($g = 1.5$ ) with $k = 1, u_0 = 1$, and $\gamma = 1$.(a) Normal dispersion ($f = 4$) (b) Anomalous dispersion ($f = -4$)

In order to see the role of the magnitude of the '$f$' parameter, we depict the contour plot of the MI gain spectrum in Fig. 3. From Fig. 3 one can observe that even on the lower branch of the dispersion curve ($f > 0$), where grating induced group velocity dispersion is normal, the system exhibits MI gain in both the above and below the threshold but not at the PT-threshold. Also, it can be seen that the gain spectrum is symmetric around $K = 0$ above the PT-threshold, while the structure exhibits MI gain only for $K < 0$ below the threshold. It is interesting to note that the MI gain decreases with increase in the value of '$g$' while opposite happens for the case above the PT-threshold. In the anomalous dispersion regime, the system shows MI gain for a small range of $K < 0$ values below the PT-threshold. The system does exhibits MI gain both at and above the PT-threshold and the gain increases with increase in the value of '$g$' parameter. It may be an interesting issue to study the gain spectrum at the top and bottom of the band gap, i.e. at $f = -1$ and $f = 1$ respectively. It can be clearly seen from Fig. 3 (c) and (d) that MI gain spectrum is distinctly different in both the cases.

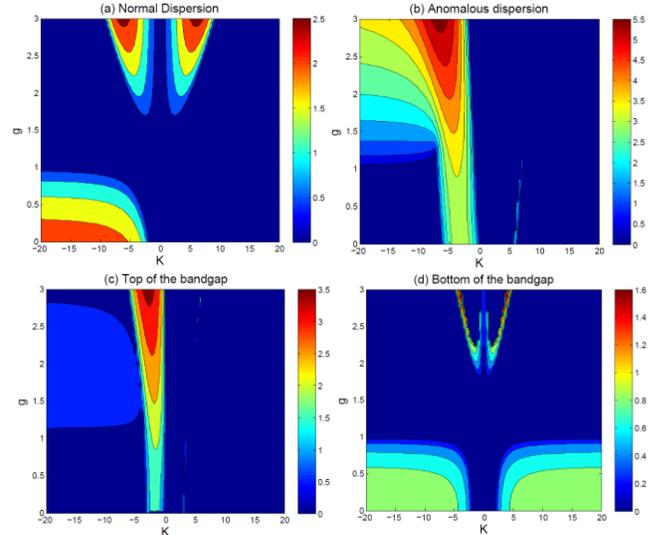

Fig.3. (Color on line) Contour plot for MI Gain with the variation of the gain/loss parameter (a) $f = 3$ (normal dispersion) (b) $f = -3$ (anomalous dispersion) (c) $f = -1$ (top of the photonic band gap) (d) $f = 1$ (bottom of the photonic band gap)

The physical origin of various features of the gain curves lie in the interaction of the four sidebands, related to the amplitudes $a_1$, $a_2$, $b_1$ and $b_2$, that are hidden in the dispersion relation. There are primarily three side-band processes. Firstly, the forward four-wave mixing (FFWM), that involves the interactions between pair of sidebands ($a_1$, $a_2$) related to the forward wave $\varepsilon_1$ or pair of sidebands ($b_1$, $b_2$) related to the backward wave $\varepsilon_2$. Secondly, the backward four-wave mixing (BFWM) that involves interaction between one forward and one backward sidebands ($a_1$, $b_2$) or ($b_1$, $a_2$). Finally, there may be interactions between the pair of sidebands ($a_1$, $b_1$) or ($a_2$, $b_2$). A rigorous but straightforward mathematical analysis in the similar line described in Ref. [39] could be carried out to understand these interactions and their role in MI.

We now briefly examine the effect of nonlinearity on the MI gain spectrum in different PT-symmetric regimes, as depicted in Fig.4. Here we consider the case of anomalous dispersion regime only.

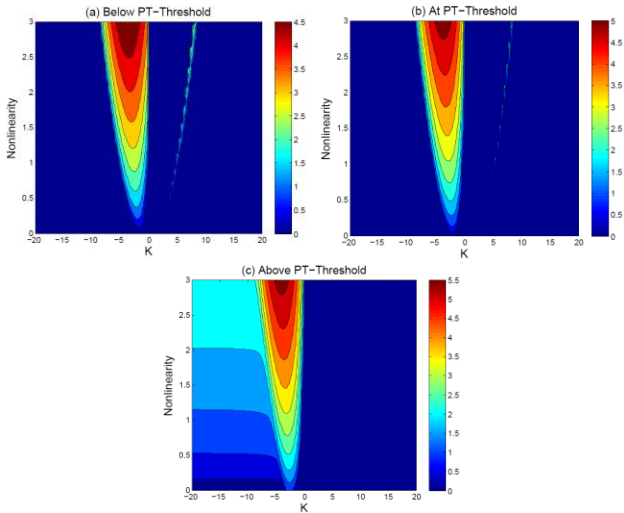

Fig.4. (Color online) Contour plot for MI Gain with the variation of nonlinear parameter, $\gamma$ with $k = 1$, $u_0 = 1$ and $f = -2$ (a) $g = 0.5$ (b) $g = 1$ (c) $g = 1.5$

One common feature in the MI gain spectrum is that, with the increase in the magnitude of the '$\gamma$' parameter, the MI-gain increases. We depict the corresponding case of normal dispersion regime in Fig. 5.

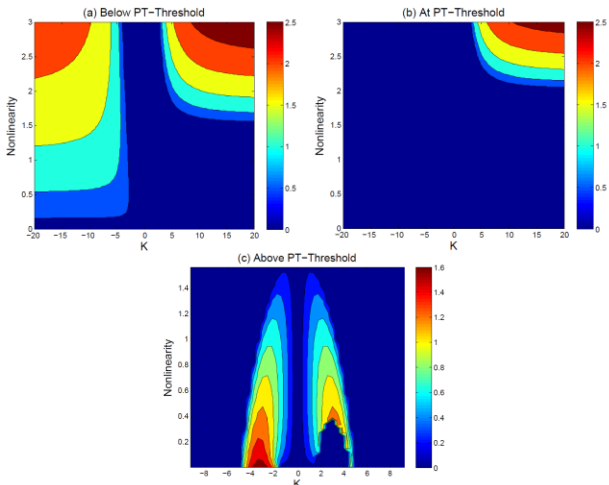

Fig.5 (Color online) Contour plot for MI Gain with the variation of nonlinear parameter, $\gamma$ with $k = 1$, $u_0 = 1$ and $f = 2$ (a) $g = 0.5$ (b) $g = 1$ (c) $g = 1.5$

An interesting feature occurs in the case of normal dispersion. We find that the structure does exhibits MI gain even at the PT-symmetry threshold, above a critical value of the nonlinear parameter. Also, above the PT-symmetry threshold the gain spectrum becomes symmetric around $K = 0$ but only above a certain value of the nonlinear parameter. In fact, if the nonlinear parameter, $\gamma < 0.4$ the MI gain plot do not exhibit symmetry between $2 < K < 4$. The origin of this behavior may be traced back to the complicated interactions among the sidebands. The perturbation frequencies, $\Omega$, take real values in the specified regime leading to no modulation instability gain.

The above analysis clearly shows that MI in such novel PT-symmetric structures is critically dependent on the PT-symmetry threshold which leaves lots of rooms for studying other associated phenomena particularly soliton formation. This study may motivate the researchers to examine all the regimes carefully. It is worthwhile to mention that our analysis is in conformity with recent results on solitary wave solutions in the nonlinear PT-symmetric Bragg grating structure [19,35]. Solitary waves are predicted in both the below and the above the PT-threshold in such settings. However, to the best of our knowledge no study has been reported regarding the solitary wave solutions exactly at the PT-symmetry threshold.

## 4.Conclusions

To conclude, we have carried out a modulation instability analysis in a nonlinear complex parity-time (PT) symmetric periodic structures. We have considered all the three regimes of PT-symmetry, namely below threshold, at threshold and above threshold. We find that at the PT-symmetry breaking point, the system do not exhibit MI in the case of normal dispersion but do show MI in the anomalous dispersion regime. This study may open up possibilities of exploiting MI in PT-synthetic periodic structures for various applications. As MI exists beyond the PT-symmetric breaking point, one may consider it to be a signature for the existence of solitons or solitary waves.

### ACKNOWLEDGMENTS
The author thanks M.A. Miri for useful discussions. This work is supported by the DST, Government of India (Grant No. SB/FTP/PS-047/2013).

### REFERENCES
1. C. M. Bender and S. Boettcher, "Real spectra in non-Hermitian Hamiltonians having PT-Symmetry" Phys. Rev. Lett. **80**, 5243-5246 (1998).
2. C. M. Bender, S. Boettcher, and P. N. Meisinger, "PT-symmetric quantum mechanics" J. Math. Phys. **40**, 2201-2229 (1999).
3. C. M. Bender, D. C. Brody, and H. F. Jones," Complex extension of quantum mechanics" Phys. Rev. Lett. **89**,270401 (2002).
4. C. M. Bender," Making sense of non-Hermitian Hamiltonians" Rep. Prog. Phys. **70**, 947-1018 (2007).
5. A. Mostafazadeh," Exact $PT$-symmetry is equivalent to Hermiticity" J. Phys. A. **36**, 7081-7091 (2003).
6. C. E. Ruter, K. G. Makris, R. El-Ganainy, D. N. Christodoulides, M. Segev, and D. Kip, "Observation of parity–time symmetry in optics", Nat. Phys. **6**, 192-195 (2010).
7. A. Regensburger, C. Bersch, M.-A. Miri, G. Onishchukov, D. N.Christodoulides, and U. Peschel," Parity–time synthetic photonic lattices" Nature **488**, 167-171 (2012).
8. L. Feng, Y. L. Xu, W. S. Fegadolli, M. H. Lu, J. E. Oliveira, V. R. Almeida, Y. F. Chen, and A. Scherer, "Experimental demonstration of a unidirectional reflectionless parity-time metamaterial at optical frequencies," Nature materials **12**, 108-113 (2013).

9. J. Schindler, A. Li, M. C. Zheng, F. M. Ellis, and T. Kottos, "Experimental study of active LRC circuits with PT symmetries," Phys. Rev. A **84**,040101(R)-1–5 (2011).
10. K. G. Makris, R. El-Ganainy, D. N. Christodoulides, and Z. H. Musslimani, "PT-symmetric periodic optical potentials", In. J. Theor. Phys. **50**, 1019-1041 (2011).
11. K. G. Makris, R. El-Ganainy, D. N. Christodoulides, and Z. H. Musslimani," Beam dynamics in PT-symmetric optical lattices" Phys. Rev. Lett. **100**, 103904 (2008).
12. Z. Lin, H. Ramezani, T. Eichelkraut, T. Kottos, H. Cao, and D. N. Christodoulides," Unidirectional Invisibility Induced by PT-Symmetric Periodic Structures" Phys. Rev. Lett. **106**, 213901 (2011).
13. L. Feng et al.," Nonreciprocal Light Propagation in a Silicon Photonic Circuit" Science **333**, 729-733 (2011).
14. T. Kottos," Optical physics: broken symmetry makes light work", Nat. Phys. **6**, 166 (2010).
15. M. A. Miri, P. L. Wa, and D. N. Christodoulides," Large area single-mode parity–time-symmetric laser amplifiers" Opt. Lett. **37**, 764-766 (2012).
16. Y. D. Chong, L. Ge and A. D. Stone," PT-symmetry breaking and laser-absorber modes in optical scattering systems" Phys. Rev. Lett. **106**, 093902 (2011); S. Longhi," PT-symmetric laser absorber" Phys. Rev. A **82**, 031801 (2010).
17. Z. H. Musslimani, K. G. Makris, R. El-Ganainy, and D. N.Christodoulides," Optical solitons in PT periodic potentials" Phys. Rev. Lett. **100**, 030402 (2008).
18. M. Nazari, F. Nazari, and M. K. Moravvej-Farshi," Dynamic behavior of spatial solitons propagating along Scarf II parity–time symmetric cells", J. Opt. Sc. Am. B. **29**, 3057-3062 (2012).
19. M. A. Miri, A. B. Aceves, T. Kottos, V. Kovanis, and D. N. Christodoulides," Bragg solitons in nonlinear - symmetric periodic potentials" Phys. Rev. A **86**, 033801 (2012).
20. V. E. Zakharov and L. A. Ostrosvsky," Modulation instability: The beginning" Physica D **238**, 540-548 (2009).
21. G. P. Agrawal, *Nonlinear Fiber Optics*, 4th Ed. (Academic Press, San Diego, 2007).
22. A. K. Sarma and M. Saha," Modulational instability of coupled nonlinear field equations for pulse propagation in a negative index material embedded into a Kerr medium" J. Opt. Sc. Am. B. **28**, 944-948 (2011).
23. M. J. Potasek," Modulation instability in an extended nonlinear Schrödinger equation" Opt. Lett. **12**, 921-923(1987).
24. P. K. Shukla and J. J. Rasmussen," Modulational instability of short pulses in long optical fibers" Opt. Lett. **11**, 171-173 (1986).
25. A. K. Sarma," Modulational instability of few-cycle pulses in optical fibers" Eur. Phys. Lett. **92**, 24004 (2010).
26. Y. Xiang, X. Dai, S. Wen and D. Fan," Modulation instability in metamaterials with saturable nonlinearity" J. Opt. Sc. Am. B. **28**, 908-916 (2011).
27. A. K. Sarma and P. Kumar," Modulation instability of ultrashort pulses in quadratic nonlinear media beyond the slowly varying envelope approximation" Appl. Phys. B **106**, 289-293 (2012).
28. N. Akhmediev and A. Ankiewicz," Modulation instability, Fermi-Pasta-Ulam recurrence, rogue waves, nonlinear phase shift, and exact solutions of the Ablowitz-Ladik equation" Phys. Rev. E **83**, 046603 (2011).
29. E. Kengne, S. T. Chui, and W. M. Liu," Modulational instability criteria for coupled nonlinear transmission lines with dispersive elements", Phys. Rev. E **74**, 036614 (2006).
30. Z. Xu, L. Li, Z. Li, and G. Zhou, "Modulation instability and solitons on a cw background in an optical fiber with higher-order effects", Phys. Rev. E **67**, 026603 (2003).
31. M. Erkintalo, K. Hammani, B. Kibler, C. Finot, N. Akhmediev, J. M. Dudley, and G. Genty, "Higher-Order Modulation Instability in Nonlinear Fiber Optics", Phys. Rev. Lett. **107**, 253901 (2011).
32. C. M. de Sterke, "Theory of modulational instability in fiber Bragg gratings", J. Opt. Soc. Amer. B, **15**, 2660-2667 (1998).
33. R. Ganapthy, K. Senthilnathan, and K. Porsezian, "Modulational instability in a fibre and a fibre Bragg grating" J. Opt. B, Quantum Semiclass. Opt. , **6**, S436-S452 (2004).
34. G. P. Agrawal, *Applications of Nonlinear Fiber Optics*, 2nd Ed. (Academic Press, San Diego, 2007).
35. S. K. Gupta and A.K. Sarma," Solitary waves in parity-time (PT)–symmetric Bragg grating structure and the existence of optical rogue waves", Eur. Phys. Lett. **105**, 44001 (2014).
36. B. J. Eggleton, R. E. Slusher, C. M. de Sterke, P. A. Krug and J. E. Sipe, "Bragg Grating Solitons", Phys. Rev. Lett. **76**, 1627-1630 (1996).
37. D. N. Christodoulides and R. I. Joseph, "Slow Bragg solitons in nonlinear periodic structures", Phys. Rev. Lett. **62**, 1746-1749 (1989).
38. W. C. K. Mak, B. A. Malomed and P. L. Chu, "Symmetric and asymmetric solitons in linearly coupled Bragg gratings", Phys. Rev. E **69**, 066610 (2004).
39. N. M. Litchinitser, C. J. McKinstrie, C. M. de Sterke, and G. P. Agrawal," Spatiotemporal instabilities in nonlinear bulk media with Bragg gratings", J. Opt. Sc. Am. B. **18**, 45-54 (2001).